\date{}
\documentclass[showpacs,twocolumn]{revtex4}
\usepackage{graphicx,psfrag,amsmath,amssymb,amsfonts,bbm,latexsym,color,dcolumn,
epsf,graphpap}

\definecolor{red}{rgb}{1,0,0}
\definecolor{blue}{rgb}{0,0,1}
\definecolor{skyblue}{rgb}{0,0,.5}
\definecolor{green}{rgb}{0,1,0}
\definecolor{orange}{cmyk}{0,.4,1,0}

\begin{document}
\title{Geometric phases in open systems: an exact model to study how they are corrected by decoherence}

\author{Fernando C. Lombardo \footnote{lombardo@df.uba.ar}}
\author{Paula I. Villar \footnote{paula@df.uba.ar}}
\affiliation{Departamento de F\'\i sica {\it Juan Jos\'e Giambiagi}, FCEyN UBA,
Facultad de Ciencias Exactas y Naturales, Ciudad Universitaria,
Pabell\' on I, 1428 Buenos Aires, Argentina}


\begin{abstract}
We calculate the geometric phase for an open system (spin-boson
model) which interacts with an environment (ohmic or nonohmic) at
arbitrary temperature. However there have been many assumptions 
about the time scale at which the geometric phase can be measured,
there has been no reported observation of geometric
phases for mixed states under 
nonunitary evolution yet. We study not only how they are 
corrected by the presence of the different type of 
environments but estimate the corresponding times at which 
decoherence becomes effective as well. These estimations should be taken 
into account when planning experimental setups to study
 the geometric phase in the nonunitary regime, particularly important
for the application of fault-tolerant quantum computation.

\end{abstract}


\maketitle

\newcommand{\beq}{\begin{equation}}
\newcommand{\eeq}{\end{equation}}
\newcommand{\dalam}{\nabla^2-\partial_t^2}
\newcommand{\mbf}{\mathbf}
\newcommand{\itm}{\mathit}
\newcommand{\beqa}{\begin{eqnarray}}
\newcommand{\eeqa}{\end{eqnarray}}

\section{Introduction}

All real world quantum systems interact with
theirs surrounding environment to a greater or lesser extent. While
closed quantum systems are bound to have an unitary evolution in
which the system's purity is preserved and the superposition
principle can be applied, open quantum systems show a different
scenario. As they are in interaction with an environment (defined as
any degrees of freedom coupled to the system which can entangle its
states), a degradation of pure states into mixtures takes place.
These mixtures states will often turn out to be diagonal in the set
of ``pointer states" \cite{Zurek} which are selected by the crucial
help of the interaction Hamiltonian ($H_{\rm int}$). They are stable
subjected to the action of $H_{\rm int}$, i.e., the interaction
between the system and the environment will leave them unperturbed.
That's exactly what makes them a ``preferred" basis. No matter how
weak the coupling that prevents the system from being isolated, the
evolution of an open quantum system is eventually plagued by
nonunitary features like decoherence and dissipation. Decoherence, 
in particular, is a quantum effect
whereby the system loses its ability to exhibit coherent behaviour.
Nowadays, decoherence stands as a serious obstacle in quantum
information processing. 

Since the work of Berry \cite{Berry}, the notion of geometric
phases (GPs) was shown to have important consequences for quantum
systems. Berry demonstrated that quantum systems could
acquire phases that are geometric in nature. He showed that,
besides the usual dynamical phase, an additional phase that was
related to the geometry of the space state was generated during an
adiabatic evolution. The original idea, framed within the context
of adiabatic and cyclic evolutions of isolated systems, has been
generalized in various aspects. While many of this propositions
have been centered around pure states, the need to address the
issue of GPs for mixed states rapidly gained prominence fuelled by
the promise of realizing quantum logic gates under realistic
physical conditions. In that direction, in
\cite{Sjoqvist}, it was introduced an alternative definition of GPs for
nondegenerate density operators based upon quantum interferometry.
In \cite{Singh}, a kinematic description of
the mixed state GP was given and its definition extended to degenerate
density operators. Recently, in \cite{Zanardi}, GPs for an open 
quantum system were studied and, finally, it was shown that the above
apparently different approaches were related in a unifying
framework. The effect on GPs of differents types of 
decoherence sources, such as dephasing and
spontaneous decay, has been analyzed \cite{Carollos}. Likewise, it  
has been shown how to generate a geometric phase through
modifications solely on the reservoir that interacts with a small
subsystem \cite{Carollos2}. 

GPs are useful in the context of quantum computation as a tool to
achieve fault tolerance. However, practical implementations of
quantum computing are always done in the presence of decoherence.
Thus, a proper generalization of the geometric phase for unitary
evolution to that for nonunitary evolution is central in the
evaluation of the robustness of geometric quantum computation.
This generalization to nonunitary evolution has been done in
\cite{Tong}, where a functional representation of GP was proposed, 
after removing the dynamical phase from the total phase
acquired by the system under a gauge transformation.

The GP for a mixed state under nonunitary evolution 
is then defined as 
\begin{eqnarray} \Phi &=&
{\rm arg}\{\sum_k \sqrt{ \varepsilon_k (0) \varepsilon_k (\tau)}
\langle\Psi_k(0)|\Psi_k(\tau)\rangle \nonumber \\
&\times & e^{-\int_0^{\tau} dt \langle\Psi_k|
\frac{\partial}{\partial t}| {\Psi_k}\rangle}\}, \label{fasegeo}
\end{eqnarray}
where $\varepsilon_k(t)$ are the eigenvalues and
 $|\Psi_k\rangle$ the eigenstates of the reduced density matrix
$\rho_{\rm r}$ (obtained after tracing over the reservoir degrees
 of freedom). In the last definition, $\tau$ denotes a time 
after the total system completes
a cyclic evolution when it is isolated from the environment.
Taking the effect of the environment into account, the system no
longer undergoes a cyclic evolution. However, we will consider a
quasicyclic path ${\cal P}:t ~\epsilon~[0,\tau]$ with
$\tau=2 \pi/\Omega$ ($\Omega$ the system's frequency) \cite{Tong}.
 It is worth
noting that the phase in Eq.(\ref{fasegeo}) is manifestly gauge
invariant, since it only depends on the path in the state space,
and that
 this expression, even though is defined for non
degenerate mixed states, corresponds to the unitary geometric
phase in the case that the state is pure (closed system)
 \cite{{Sjoqvist},{Singh}}.

It is expected that GPs can be only observed in experiments
carried out in a time scale slow enough to ignore nonadiabatic corrections, 
but rapid enough to avoid the destruction of the interference 
pattern by decoherence \cite{Gefen}. So far, there has been no 
experimental observation of 
GPs for mixed states under nonunitary evolutions. The purpose 
of this letter is 
to study how GPs are affected by decoherence.
Not only shall we analyze the effect of the environment on the GPs
and their robustness against decoherence but also, under which
conditions GPs can be measured. With these motivations, we shall
introduce a ``simplified" spin boson model and calculate the
corrections to the unitary geometric phase for different physical
environments. In the end, we shall estimate the decoherence times
after which GPs can be no longer measured because they, literally,
disappear.

The paper is organized as follow. In Section II we introduce the spin-boson 
model and we show the exact master equation for the reduced density matrix. We 
explicetly evaluate, in Section III, the corrections to the GPs induced by 
different types of environment and Section IV contains the calculation 
of the decoherence times. Finally, in Section V, we make our final remarks and 
comments.

\section{The spin-boson model} 

Hereafter, we shall study an exact model
for a two-state quantum system [or quantum bit (qubit)] coupled to
a thermal bath of harmonic oscillators, where decoherence is the
only effect on the system-particle. This is a particular case of
the spin-boson model of Ref.\cite{Leggett} (where the tunnelling
bare matrix element is $\Delta=0$) and has been used by many
authors to model decoherence in quantum computers \cite{Eckert}
and, in particular, it is extremely relevant to the proposal 
for observing GPs 
in a superconducting nanocircuit \cite{Falci}. We will
concentrate on ohmic and supraohmic environments (at high, low,
and zero temperature) coupled to the system spin. The Hamiltonian
that describes the complete evolution of the two-state system
interacting with the external environment is:
\begin{equation}
H_{\rm SB}=  \frac{1}{2} \hbar \Omega
 \sigma_z +\frac{1}{2} \sigma_z
\sum_{n} \lambda_{n} q_{n} + \sum_{n} \hbar \omega_{n}
a_{n}^{\dag} a_{n},
\end{equation}
where the environment is described as a set of harmonic
oscillators with a linear coupling in the oscillator coordinate.
The interaction between the two-state system and the environment
is entirely represented by a Hamiltonian in which the coupling is
only through $\sigma_z$. A coupling of this form indicates that
the state of the environment will be sensitive to the values of
$\sigma_z$, which means that the environment can ``observe" the
value of $\sigma_z$ (i.e, where the system is $|\uparrow>$ or
$|\downarrow>$). The reason for this type of coupling is that the
effects of a coupling proportional to $\sigma_x$ and/or $\sigma_y$
can be completely taken into account by the renormalization of the
natural frequency of the system. However, in this particular case,
$[\sigma_z,H_{\rm int}]=0$ and the corresponding master equation is
much simplified, with no frequency renormalization and dissipation
effects. If we assume that
(i) the system and the environment are initially uncorrelated and
(ii) the environment is initially in thermal equilibrium at
temperature $T$ (or the vacuum for zero temperature), the master
equation for the reduced density matrix is
\begin{equation}
\dot{\rho_{\rm r}} = -i \Omega [\sigma_z,\rho_{\rm r}] - {\cal D}(t)
[\sigma_z,[\sigma_z,\rho_{\rm r}]]. \label{masterH_0}
\end{equation}
In other words, the model describes a purely decohering mechanism,
solely containing the diffusion term ${\cal D}(t)$, where no
energy exchange between system and bath is present. The diffusion
coefficient is given by
\begin{equation}
{\cal D}(t)=\int_0^tds\int_0^{\infty}
d\omega J(\omega) \cos(\omega s)\coth(\frac{\beta \hbar \omega}{2}),
\label{D}
\end{equation}
where $\beta=1/k_B T$ ($k_B$ is Boltzman constant) and $J(\omega)$
is the spectral density of the environment defined by the
expression $J(\omega)= \sum_n \lambda^2_n \delta(\omega-\omega_n)/2
m_n \omega_n$. One assumption we shall make throughout this paper
is that $J(\omega)$ is a reasonably smooth function of $\omega$,
and that is of the form $\omega^n$ up to some frequency $\Lambda$
that may be large compared to $\Omega$. In particular, the case with
$n=1$ is the ``ohmic" case and the one with $n>1$ is the
``supraohmic" case. It is easy to see in Eq.(\ref{D})
that the system's dynamics depends crucially
both on the external temperature and on the type of environment.

\section{Geometric phase for the composite system} 

We shall compute
the geometric phase for the spin-boson model. It is easy to
check that $\rho_{\rm r_{01}}(t)= e^{-i \Omega t -
\Gamma(t)}\rho_{\rm r_{01}}(0)$ is 
solution of the master equation (Eq.(\ref{masterH_0})) with
$\Gamma(t)=\int_0^{t}~dt' {\cal D}(t')$ and $\rho_{\rm r_{01}}(0)$ a 
constant determined by initial conditions. Here we are implying
that the diagonal terms do not evolve in time, i.e. 
$\rho_{\rm r_{ii}}(t)=\rho_{\rm r_{ii}}(0)$, for $i=0,1$ (the spin state basis),
 assuming the population dynamics of the system is
essentially frozen on the time scales of interest \cite{Eckert}. The decoherence
factor is then given by
 \begin{equation}
 \Gamma(t)= 4 \int_0^{\infty} d\omega ~ J(\omega)~
{\mathrm {coth}}\bigg(\frac{\beta \hbar
\omega}{2}\bigg)~\frac{(1-\cos(\omega t))}{\omega^2}. \label{G}
 \end{equation}
 When a specific choice turns out to be useful,
 we shall assume the following functional form for the spectral
density $J(\omega)= \gamma_0/4 \omega^n \Lambda^{n-1} e^{-\omega/\Lambda}
$ \cite{Leggett}, where $\gamma_0$ is the dissipative constant
 (in suitable units) and $\Lambda$ is the cutoff frequency.

As the factor $\Gamma$ is real, unitarity requires
 $\rho_{\rm r_{01}}(0)=\rho_{\rm r_{10}}(0)$ since $\rho_{\rm r_{10}}=
{\rho_{\rm r_{01}}}^{*}$.
For a spin $1/2$ system, the state space consists of all points on
and inside the Bloch sphere. Then, we assume that the system is
initially in the Bloch state
 
\begin{equation} 
|\Psi (0) \rangle=\cos(\theta_0/2) |e \rangle + \sin(\theta_0/2) |g \rangle .
\nonumber\end{equation}
The constants
are determined so that $\rho_{\rm r}(0)=|\Psi(0)\rangle \langle \Psi(0)|$. 
Then, for times $t>0$, the reduced density matrix is
\begin{equation}
\rho_{\rm r}(t)=\bigg(\begin{array}{cc} \cos(\theta_0/2)^2 ~~~~~~~~~~~~
 1/2 \sin(\theta_0) e^{i \Omega t - \Gamma(t)}\\
1/2 \sin(\theta_0)e^{-i \Omega t - \Gamma(t)}~~~~~~~~~~~~
 \sin(\theta_0/2)^2
\end{array} \bigg). \nonumber
\end{equation}

The eigenvalues of the above reduced density matrix
 are easily calculated:
\begin{equation}
\label{varep+}
\varepsilon_{\pm}(t) = \frac{1}{2} \pm \frac{1}{2}
\sqrt{\cos(\theta_0)^2
+ \exp(-2 \Gamma (t)) \sin(\theta_0)^2}. 
\end{equation}
In order to estimate the geometric phase, we only need the
eigenvector $|\Psi_{+}(t) \rangle$ since $\varepsilon_{-}(0)=0$,
and, hence, the only contribution to the phase comes from that
eigenvector and its corresponding eigenvalue (see
Eq.(\ref{fasegeo})). Then, we write it here as:
\begin{equation}
\label{Phi+}
|\Psi_{+} (t) \rangle = e^{-i \Omega t} \sin(\theta_t/2)
|e\rangle + \cos(\theta_t/2) |g\rangle, 
\end{equation}
with $\tan (\theta_t/2) = \exp(- \Gamma(t) ) \cot (\theta_0/2)$.
It is easy to check that, when $\Gamma=0$, we reobtain the results
for the unitary case. Hence, once we have the eigenvalues and eigenvectors, we can
calculate the factor $\langle \Psi_k| \dot{\Psi}_{k} \rangle$ of
Eq. (\ref{fasegeo}). Performing the time derivatives, we get
$\langle \Psi_k| \dot{\Psi}_{k} \rangle= -i \Omega
\sin(\theta_t/2)^2$, therefore, the geometric phase is
\begin{equation}
\Phi={\mathrm {arg}}\bigg\{\sqrt{\varepsilon_+(\tau)
\varepsilon_+(0)} \langle \Psi_+(\tau)| \Psi_+(0)\rangle e^{i
\Omega \int_0^{\tau}dt  \sin(\frac{\theta_t}{2})^2}\bigg\}.
\nonumber
\end{equation}
This result holds for any density matrix has these eigenvalues and
eigenvectors, independently of the exact expression of $\Gamma$,
as long as $\varepsilon_-(0)=0$. By inserting Eqs. (\ref{varep+})
and (\ref{Phi+}) into the above expression, the geometric
phase related to a quasicyclic path ${\cal P}:t ~\epsilon~[0,\tau]$ (with
$\tau=2 \pi/\Omega$  \cite{Tong}) is
 (assuming $\cos(\theta_0/2) \geq 0$):
\begin{equation}
\Phi= \Omega \int_0^{\tau}~dt~ \sin
\bigg(\frac{\theta_t}{2}\bigg)^2.
\end{equation}

In order to evaluate this integral, we will perform
 a serial expansion in terms of powers of the dissipative
 constant $\gamma_0$, and,
 consequently, the unitary phase $\Phi^{U}$
  is corrected
 by the presence of the environment as:
\begin{eqnarray}
\Phi&=& \Phi^{U} + \delta \Phi
\approx  \pi (1-\cos(\theta_0)) \nonumber \\
&+& \gamma_0 \frac{\Omega}{2} \sin(\theta_0)^2 \cos(\theta_0)
\int_0^{\tau}~dt \bigg[\frac{\partial
\Gamma(t)}{\partial \gamma_0} \bigg]\bigg|_{\gamma_0=0}, \nonumber
\end{eqnarray}
where we can see that the first term of the expansion is the
solution we would have obtained if the evolution would have been
unitary, i.e $\Phi^{U}=\pi (1-\cos(\theta_0))$.
If, for example, we assume an ohmic ($n = 1$) environment at high
temperature, then the spectral density takes the particular form
$J(\omega)= \gamma_0/4~ \omega e^{-\omega/\Lambda}$. In that case,
$\hbar \omega << 2 k_B T$, we can
approximate ${\mathrm {coth}}(\beta \hbar \omega /2)$ in Eq. (\ref{D})
by $2 k_B T/(\hbar \omega)$ and the decoherence coefficient $\Gamma$
becomes $\Gamma= (\gamma_0 \pi k_B T) t/\hbar$. Then, the geometric
phase is corrected by:
\begin{equation}
\delta \Phi = \pi^2 \gamma_0 \bigg(\frac{\pi k_B T}{\hbar
\Omega} \bigg) \sin(\theta_0)^2 \cos(\theta_0). \label{ohHT}
\end{equation}
This correction to the unitary phase is proportional to 
$\gamma_0 k_B T/\hbar \Omega$. Then, it is bigger for hotter environments
and can not be neglected. It is worth mentioning that, formally, our
solution is similar to the one proposed by authors in \cite{Yi}
where they made no a priori assumption about the dynamics of the
system. On the contrary, our solution is obtained from scratch and
can be applied to different environments at any temperature. What's
more, this correction is more realistic since they assumed a 
very much simplified environment.

If we were to assume the same ohmic environment but at zero
temperature, then the phase would be corrected in a significantly
different way, as one would expect since there is one time scale
lost. In that case, the factor ${\mathrm {coth}}(\beta\hbar\omega
/2)$, in the definition of $\Gamma(t)$, (Eq. (\ref{D})) can be
approximated by $1$ and the correction obtained is:
\begin{equation}
\delta \Phi= \pi \gamma_0 \bigg(-1+\log(\frac{2 \pi
 \Lambda}{\Omega})\bigg)
\sin(\theta_0)^2 \cos(\theta_0). \label{ohT0}
\end{equation}
It is worthy noting that this correction comes from the zero point
fluctuations of the environment.

Beyond the commonly assumed ohmic spectrum for the bath, generic 
nonohmic environments can be studied with this model. Owing to the
ultrashort time bath correlations, nontrivial short-time system
dynamics enters. What's most appealing about this case is the fact
that the electromagnetic field can be modeled by a supraohmic
environment which 
results very useful for quantum optics and trying
to measure the geometric phases. In the following, we shall evaluate
 the correction of the geometric phases in the
presence of this kind of environments.

For high temperature, we can still make the same approximation
than in the ohmic case, and then, the estimation of the
$\Gamma$ factor is straightforward. Using the spectral density 
with $n=3$, the correction to the unitary phase is:
\begin{equation}
\delta \Phi= \pi \gamma_0 \bigg( \frac{2 k_B T}{\hbar
\Lambda} \bigg) \sin(\theta_0)^2 \cos(\theta_0).\label{nohHT}
\end{equation}
We can also estimate the correction to the unitary phase in the
presence of an supraohmic environment but at zero temperature as:
\begin{equation}
\delta \Phi = \pi \gamma_0
 \sin(\theta_0)^2 \cos(\theta_0).
 \label{nohT0}
\end{equation}
As $\Lambda$ is the maximum available
frequency in the environment, a valid assumption is that $\Omega
\lesssim \Lambda$. In that case, we can see that the parenthesis in
Eq.(\ref{ohT0}) is of order one and, then, both corrections at
zero temperature (Eq.(\ref{ohT0}) and Eq.(\ref{nohT0})), for the
ohmic and nonohmic case, are similar. The same occurs for high
temperate in Eq.(\ref{ohHT}) and Eq.(\ref{nohHT}). 
These results are remarkably interesting and
enhance the robustness of the model (and of the geometric phase).

\section{Decoherence times for the composite system} 

In this section we
shall estimate the decoherence time, i.e. the
time-scale when the nondiagonal terms of the reduced density
matrix are suppressed. Therefore,
it is a very important time scale to take into account when
planning an experiment to measure the geometric phases.

Let's take for example an interference experiment. The experiment
starts by the preparation of two wave packets in a
coherent superposition, assuming each of the particles
follows a well defined classical path ($C_1$ and $C_2$,
respectively), as $\Psi(t=0)=(\varphi_1(x)+\varphi_2(x))
\otimes \chi_0(y)$,
where $\chi_0(\vec y)$ represents the initial quantum state of the
environment (whose set of coordinates is denoted by $\vec y$). Due
to the interaction between the system and the environment, the
total wave function at a later time $t$ is
$\Psi (t) =  \varphi_1(\vec x,t) \otimes \chi_1 (\vec y,t) +
\varphi_2(\vec x,t) \otimes \chi_2 (\vec y,t)$. 
It is easy to note that the states  $\varphi_1$ and
$\varphi_2$ became entangled with two different states of the
environment. Therefore, the density matrix of the system at time
$t$ (for example when interference pattern is examined) is,
\begin{eqnarray}
\rho_{\rm r}(x,x') &=& \varphi^{*}_1(x) \varphi_1(x') +  \varphi^{*}_2 (x)
\varphi_2(x') \nonumber \\
&+& \bigg(\varphi^{*}_1(x) \varphi_2(x')  + \varphi^{*}_2(x)
\varphi_1(x')\bigg) {\cal F}(t). \nonumber \label{density}
\end{eqnarray}

The last two terms in the above expression represent the quantum
interferences. The overlap factor ${\cal F}(t)$ 
\cite{{PRA}} encodes the
information about the statistical nature of noise since it is
obtained by tracing over the degrees of freedom of the bath.
Hence, it produces a decaying term that
tends to eliminate the interference pattern. 
\begin{figure}[t]
\begin{center}
\includegraphics[width=8.6cm]{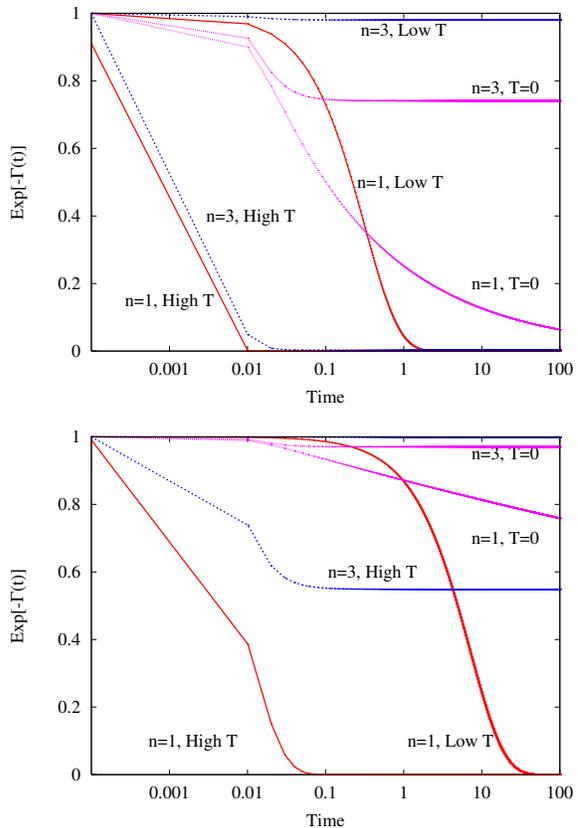}
\caption{Comparison between the $\Gamma(t)$ coefficients for the
ohmic case ($n=1$) and the supraohmic case ($n=3$) for different
environments for high ($T=1000$), low ($T=1.55$) and
zero temperature ($\hbar=1=k_B)$. $\Lambda=100\Omega$ and 
$\gamma_0=0.3$ (top) and $\gamma_0=0.03$ (bottom). Time is 
measured in units of $\Omega$. It is easy to note that the 
low temperature limit is not zero temperature.}
\label{coefi}
\end{center}
\end{figure}
Therefore, noise makes ${\cal F}(t)$ less than one, and the goal is to
quantify how it slightly destroys the particle interference
pattern. In the case of the exact system we are studying here,
${\cal F}(t)=e^{-\Gamma (t)}$. When $\Gamma$ is big, the off-diagonal terms
(coherences) elements will vanish in a short time scale, and hence,
the open system will not acquire GP. Then, in order to know 
the time-scale at which the quantum
interferences are suppressed, we have to estimate the
``decoherence" time $t_D$ for each of the four cases studied above,
by setting $\Gamma (t_D) \approx 1$.

For the ohmic environment, the decoherence factor is 
$\Gamma^{1T}=(k_B T \pi \gamma_0/\hbar) t$ 
in the limit of high temperature.
Then, the decoherence time scale for this case is, in principle, 
extremely short, estimated as $t_D^{1T}=\hbar/(k_B T \pi \gamma_0)$ (see
Fig.\ref{coefi}).  All other cases
depend explicitly upon the cutoff.  
For the ohmic
environment at zero temperature, the decoherence factor is 
$\Gamma^{10}=\gamma_0/2 \log(1+\Lambda^2 t^2)$. 
For times $\Lambda t \geq 1$, the decoherence time scales as 
$t_D^{10}= e^{1/\gamma_0}/\Lambda$. 
Sooner or later, decoherence is always 
present in this environment as well. As it is clearly shown
in Fig.\ref{coefi}, decoherence is delayed as $\gamma_0$ 
decreases (bottom).
Things are slightly different for the supraohmic environment.
In the high temperature limit, the decoherence factor is 
$\Gamma^{3 T}=(2 k_B T \gamma_0) \Lambda t^2/\hbar$ for 
$\Lambda t \ll 1$. Then, the estimation of the decoherence 
time is straightforward  $t_D^{3 T}=1/\Lambda 
\sqrt{\hbar \Lambda/(2 k_B T \gamma_0)}$. However, if $\Lambda t \geq 1$, 
$\Gamma^{3 T}= 2 k_B T \gamma_0/(\hbar \Lambda)$,
constant in time. Then, decoherence shall occur if and 
only if $2 k_B T \gamma_0 \gg \hbar \Lambda$, and that shall 
happen in a time $t < 1/\Lambda$ as 
it is shown on top of Fig.\ref{coefi}. At the bottom of 
Fig.\ref{coefi}, $\gamma_0$ is
smaller and then, $k_B T \gamma_0 < \hbar \Lambda$, which means
that $\Gamma^{3 T}$ will never be of order one and, therefore,
decoherence shall not be effective. Finally, in the
supraohmic case at zero temperature, the decoherence factor
is $\Gamma^{3 0}= \gamma_0 \Lambda^4 t^4/(1+\Lambda^2 t^2)^2$. 
We can see in Fig.\ref{coefi}
that decoherence never occurs for this case. This is so because
$\Gamma^{3 0}= \gamma_0$ for $\Lambda t \geq 1$. The
decoherence factor will be a constant value of $e^{-\gamma_0}$ for 
all times. Since $\gamma_0 < 1$, this factor will
never be of order one. In conclusion, as high-T environments are 
so efficient producing decoherence, 
there is a strong condition over the dissipative constant $\gamma_0$ in order 
to get $t_D > \tau$ \cite{Zanardi2}, i.e. $\gamma_0 < \hbar\Lambda/k_BT$, which  
actually is unfeasible. Bigger $\gamma_0$ implies shorter decoherence times. 
However, for zero-T baths, the condition on $\gamma_0$ relaxes 
to $\gamma_0 < 1$ making $t_D > \tau$ and allowing  
experimental observation. 

\section{Final remarks}

When measuring a GP, the dynamical phase has to 
be eliminated.
It can be either canceled, for example, using spin-echo technique
for spins in magnetic fields \cite{Ekert}, or one can parallel transport the
state vector in order to ensure that the dynamical phase is zero 
at all times. Even though the state does not acquire a phase 
locally, it can acquire
a phase globally after completing a cyclic evolution. This 
global phase is equal to the geometric phase. GPs 
are useful in the context of quantum computation as a tool 
to achieve fault tolerance. However, practical 
implementations of quantum computing are always
done in the presence of decoherence, a nonunitary 
effect of open systems.
However, there have been many assumptions about the 
time scale at which the geometric phase can be measured, 
there 
has been no reported observation yet for mixed states 
under nonunitary evolutions. 

It has been argued that
the observation of GPs should be done in times long enough to
 obey the adiabatic approximation but short enough to prevent
decoherence from deleting all phase information. In this article, 
not only have we shown an exact model where the correction 
to the GPs can be evaluated for different types of environment 
at any temperature but also, estimations of the
corresponding times at which the interference pattern is mostly 
reduced and decoherence becomes effective. 
This model allows us to evaluate an exact master equation for the 
reduced density matrix and, as a consequence, to elaborate a full analysis 
on the effect of decoherence on the geometric phase of the composite system.
The goal is the analysis of the effect produced by decoherence, which should 
be essential to design experimental setups in order to observe geometric phases 
using, for example, interferometry. 
We stressed that the convenient environment to observe GPs is at zero
temperature in the underdamped limit and that these decoherence times 
should be taken into account when
extending the study of GPs to the nonunitary regime, especially
pertinent to the application to fault-tolerant quantum computation
\cite{Du}.

\section{Acknowledgments}

We thanks F. Mazzitelli for useful comments. 
This work was supported by UBA, CONICET, 
and ANPCyT, Argentina.

\end{document}